\documentclass[mathleft
]{an}
\usepackage{graphicx}
\usepackage{times}
\begin{document}

\Pagespan{1}{}
\Yearpublication{}%
\Yearsubmission{}%
\Month{}%
\Volume{}%
\Issue{}%

\title{Towards an Accurate Model for the Antennae Galaxies}

\author{S.J. Karl\inst{1}\fnmsep\thanks{\email{skarl@usm.uni-muenchen.de}\newline}
\and  T. Naab\inst{1}
\and P.H. Johansson\inst{1}
\and Ch. Theis\inst{2}
\and C.M. Boily\inst{3}
}
\titlerunning{Towards an Accurate Model for the Antennae}
\authorrunning{Karl et al.}
\institute{
Universit\"{a}ts-Sternwarte M\"{u}nchen, Scheinerstr. 1, 
D-81679 M\"{u}nchen, Germany
\and
Institut f\"{u}r Astronomie der Universit\"{a}t Wien,
T\"{u}rkenschanzstr. 17, A-1180 Vienna, Austria
\and
Observatoire astronomique, Universit\'e de Strasbourg and CNRS UMR 7550,
11 rue de l'Universit\'e, F-67000 Strasbourg, France
}

\received{}
\accepted{}
\publonline{}


\abstract{
In the framework of hierarchical structure formation
ellipticals can form from merging of smaller disk
galaxies. The nearby interacting 'Antennae' galaxy
pair (NGC 4038/39) is one of the best-studied local
systems of merging spirals, thus presenting us with an ideal
laboratory for the study of galaxy evolution models. The
Antennae are believed to be in a state prior to their
final encounter with rapid subsequent merging,
which puts them in the first position of the
Toomre (1977) merger sequence. Here we present first numerical high-resolution,
self-consistent, smoothed particle hydrodynamics (SPH)
simulations of the Antennae system, including star formation and
stellar feedback, and compare our results to VLA HI observations by
Hibbard et al.~(2001). We are able to obtain a close, but not yet
perfect match to the observed morphology and kinematics of the
system.
}

\maketitle
\section{Introduction}
\begin{figure}
\centering
\includegraphics[width=50mm,height=41.25mm,angle=90, scale=1.]{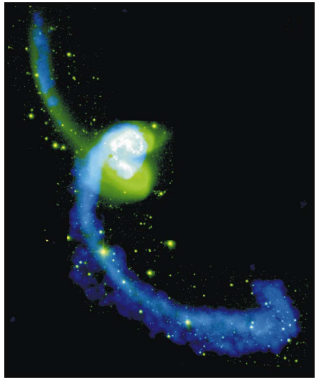}
\caption{NGC 4038 (south) and NGC 4039 (north). Distribution of HI and optical light. HI data are shown in blue, together with a combined B + V + R optical image in
  white and green. Taken from Hibbard et al.~(2001). North is pointing
upwards.}
\label{FigHibbHIoptical}
\end{figure}

The 'Antennae' galaxies (NGC 4038/39, Arp 244, VV245) are a
well-known, archetypal example of a spiral-spiral\linebreak galaxy merger (see
Fig. \ref{FigHibbHIoptical}). They are
nicknamed after their spectacular appearance with a prominent pair of tidal
tails, which was already noted in early
optical images (e.g. Duncan 1923, Schweizer 1978).
At present there is a huge\linebreak amount of data collected for the Antennae both from\linebreak ground- and space-based
telescopes, covering a wide range of wavelength regimes. Recent examples include:
optical HST WFPC data (Whitmore et al.~1999), near-IR WIRC (Brandl et al.~2005) and
mid-IR IRAC (Wang et al.~2004) imaging, and Chandra ACIS-S X-ray observations (Baldi et
al. 2006). The elongated tails are a clear sign of tidal
interaction of nearly equal-mass galaxies (Toomre \& Toomre
1972). They form kinematically by gravitational tides exerted during
the interaction process and their
morphology and velocity fields give strong hints on the
encounter geometry and history.
This makes the Antennae galaxies a key system for investigating galaxy interactions and
their associated physical phenomena. Starting from the first numerical
model by Toomre \& Toomre (1972), where they used the restricted
N-body method to model the evolution of the disks, there have been many
attempts to obtain numerical 'look-alikes' for the Antennae galaxies. Barnes
(1988) was the first to use self-consistent N-body models with
multiple components consisting of a bulge, a disk, and dark
halo with a mass ratio of 1:3:16 and a total mass of $2.75 \cdot
10^{11}\, \mathrm{M}_\odot$. Dubinski, Mihos \& Hernquist~(1996) used
the extent of the tidal features in
the Antennae to probe the amount of dark matter in these
galaxies. Mihos, Bothun \& Richstone~(1993) were the first to include gas dynamics and
star formation into their dynamical model of the Antennae
galaxies. For the latest review of the Antennae models we would also
like to refer to the results presented by Hibbard~(2003). However, most
of the Antennae models are still based on the orbital parameters given
in Toomre \& Toomre (1972). Here we want to extend the probed
parameter space in order to find orbital parameters which allow for a closer fit to the
observed morphological and
kinematical data of the Antennae. We use medium-resolution VLA HI mappings of the Antennae ($\sim
20$'', $\Delta v=5.21 \mathrm{km}\,\mathrm{s}^{-1}$)
by Hibbard et al.~(2001) for comparison with our models. In this case it is
advantageous to use the cold atomic gas as
a sensitive tracer of the large-scale dynamics of the system as it is unlikely to be disturbed by star formation. Recently,
there has been a debate about the exact distance to the Antennae,
ranging from a modest $13.3 \pm 1.0$ Mpc (Saviane et al. 2008), based on
photometry of the tip of the red giant branch, to $22 \pm 3$ Mpc
(Schweizer et al.~2008), based on observations of a supernova of type
Ia. Sometimes even higher and lower values have been quoted
(e.g. Zezas \& Fabbiano 2002, Rubin, Ford \& D'Odorico 1970). A
widely-used, intermediate distance to the Antennae is $D = 19.2$ Mpc,
which is derived from the systemic recession velocity relative to the
Local Group assuming a Hubble constant of $H_0 = 75
~\mathrm{km}~\mathrm{s}^{-1}~\mathrm{Mpc}^{-1}$ (Whitmore et
al.~1999). In this paper the distance will be adopted as part of the
model matching process.
\begin{figure}
\centering
\includegraphics[width=50mm,height=50mm,scale=1.]{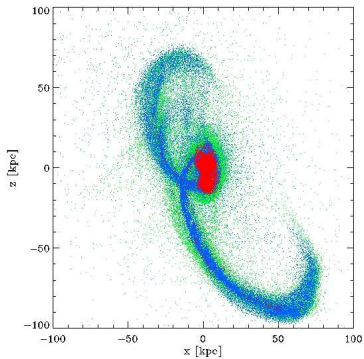}
\caption{'Best match' to the Antennae at $t=0.60$ Gyr in the plane-of
  the sky. Disk and bulge particles are shown in green,
  gas in blue, and stars, which have formed during the simulation, in red.
  Note that for the old stellar component (green) we plot only every
  fifth particle.}
\label{FigPlaneOfSky}
\end{figure}
\section{Numerical setup}
For our high-resolution model of the Antennae galaxies we set up
equilibrium galaxy models following Springel, di\linebreak Matteo \&
Hernquist~(2005). Each model consists
of a NFW halo (Navarro, Frenk \& White 1997)
 which is then converted to a Hernquist (1990) profile dark matter
 halo. Embedded in the dark
 halo is a stellar Hernquist (1990) bulge component, and an additional
 exponential stellar and\linebreak gaseous disk. The ratio
of luminous to dark matter, or, the baryon mass fraction, is $f_b = M_\mathrm{bary}/M_\mathrm{DM}= 1/4$, where
NGC 4038 is modeled as an 'Sc-type galaxy' with a bulge-to-disk
ratio of B/D = 0.2 and NGC 4039 as an 'Sb-type galaxy' with a 
B/D = 0.4. Each galaxy has a total mass of $M_\mathrm{tot} = 2.77\cdot 
10^{11}\, \mathrm{M}_\odot$. As a starting point for our models we adopted the baryon mass fraction and the
total masses from Barnes (1988). The initial
gas fraction in the disks was chosen to be 20\%, which
amounts to $M_\mathrm{gas}^{4038} \simeq 9.2 \cdot 
10^{9}\, \mathrm{M}_\odot $ and $M_\mathrm{gas}^{4039} \simeq 7.9 \cdot 
10^{9}\, \mathrm{M}_\odot $, respectively. Further parameters
for the two model galaxies are summarized in Tab. \ref{tab1}. 
All simulations are run on the local ALTIX SGI supercomputer using the fully parallel smoothed particle
hydrodynamics (SPH) (see e.g. Monaghan 1992) code Gadget2 (Springel
2005). The simulations include a prescription of radiative cooling for
primordial hydrogen and helium (Katz, Weinberg \& Hernquist 1996). We also model star
formation and supernova feedback, following the sub-grid multiphase
prescription as described by
Springel \& Hernquist (2003), but exclude supernovae-driven galactic winds. 
Initially the galaxies move on elliptical Keplerian orbits ($e \approx 0.8$) 
with a pericentric separation $r_\mathrm{p} = 20$ kpc and
an initial separation of one virial radius $r_\mathrm{init} = 106$
kpc. We ran a set of low-resolution simulations, going through
iterative cycles, while varying the scaling parameters, the orientation of
the disks and the viewing angle, until we obtained a good match with the observational
data. However, this resulted in a rather high adopted distance to the
Antennae of $D = 32.6$ Mpc if we choose to apply no spatial scaling. This merger geometry was then re-simulated at high-resolution including star
formation and radiative cooling. The total number of particles was
$N_\mathrm{tot} =$ 1,600,000, of which NGC 4038 and NGC4039
contributed 105,000 and 180,000 bulge particles, 420,000 and 360,000
disk particles, and 105,000 and 90,000 gas particles,
respectively. Furthermore, each galaxy
consisted of $N_\mathrm{DM} = 170,000$ dark matter halo particles. The particle numbers were chosen in this particular way
for dynamical reasons, i.e. to ensure that all particles in the baryonic
component have exactly the same masses in order to minimize two-body
relaxation effects. The
gravitational softening parameters for stellar and gaseous particles
were set to $\epsilon = 0.020$ kpc and for the dark matter halo
particles to $\epsilon = 0.083$ kpc, and the system was evolved
for a total time of $\sim 2$ Gyr.

\begin{table}
 \centering
\caption{Galaxy model parameters}
\label{tab1}
\begin{tabular}{ccc}\hline
Modeled Property & NGC 4038 & NGC 4039\\ 
\hline
Disk scale length & 3.31 kpc & 3.29 kpc\\
Disk scale height & 0.66 kpc & 0.66 kpc\\
Bulge scale length & 0.66 kpc & 0.66 kpc\\
Maximum rotational velocity & 169 km/s & 170 km/s\\
\hline
\end{tabular}
\end{table}
\section{Results}
\begin{figure}
\centering
\includegraphics[width=45mm,height=135mm]{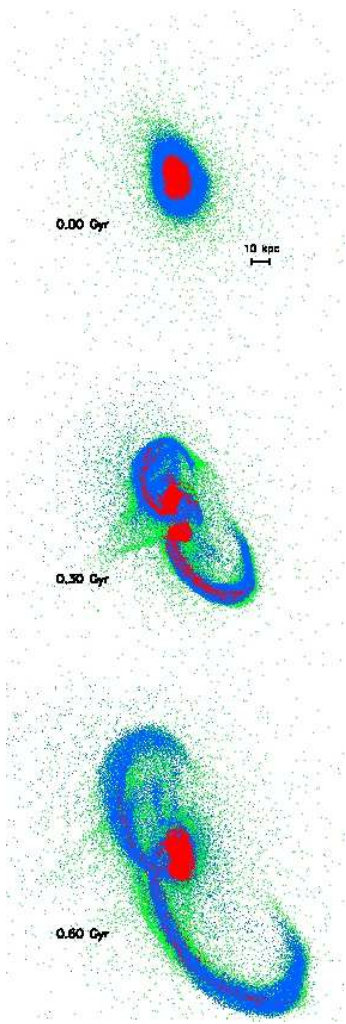}
\caption{Dynamical evolution in the NGC 4038/39 simulations. We show a
  sequence of snapshots at three equally-spaced times between
  pericenter ($t = 0.0$ Gyr) and best match ($t = 0.82$ Gyr). Colors are coded the same way as
 in Fig. \ref{FigPlaneOfSky}}
\label{FigMergerTimeSequence}
\end{figure}

\begin{figure*}
\centering
\includegraphics[width=171mm,height=53mm]{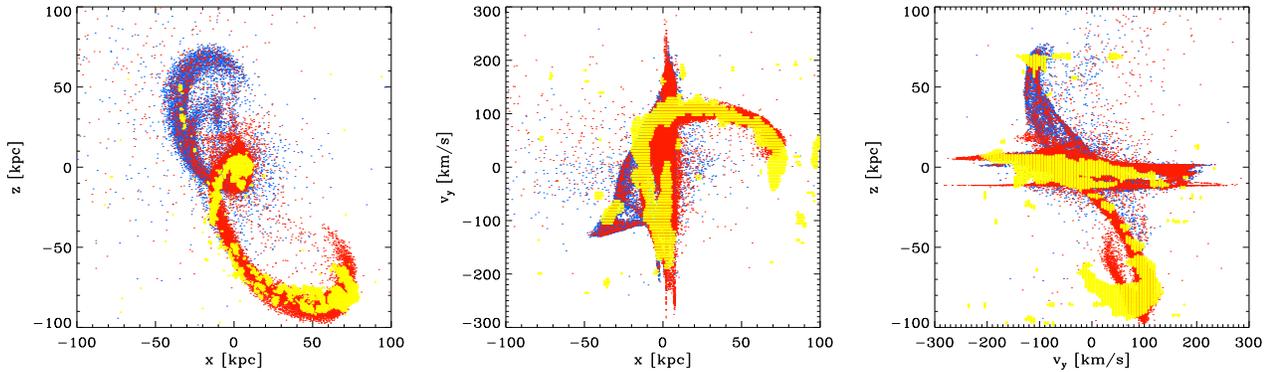}
\caption{Position-velocity diagram of the Antennae model at time of
  'best fit': X-Z (plane of the sky,
  left), X-V$_\mathrm{y}$ (middle), and V$_\mathrm{y}$-Z
  (right). Only gas particles for both simulated galaxies NGC 4038 (red) and NGC
  4039 (blue) are shown. HI data (Hibbard et al.~2001) are overlaid in yellow.}
\label{FigVcube}
\end{figure*}

Fig. \ref{FigMergerTimeSequence} shows the evolution of the
Antennae merger morphology as a function of time. In our simulation,
we set the origin of the time such that the galaxies pass pericenter
at a time $t = 0$ Gyr (upper panel in Fig. \ref{FigMergerTimeSequence}), and have
their final encounter at $t \sim 0.82$ Gyr.
Star formation is mostly
confined to the central regions of the simulated galaxies.
At the time we stopped our calculations, $t \simeq 1.4$ Gyr, the
galaxies still show elongated tidal tails and clear signs of a tidally disturbed morphology
at their centers. They have not yet transformed into an elliptical-like merger
remnant (see, e.g. Naab \& Burkert 2003, Naab, Jesseit \& Burkert
2006, Johansson, Naab \& Burkert~2008). We
obtain our 'best fit' at a time shortly before the second encounter,
that is $t \approx$ 600 Myr
after pericenter and $t \approx$ 220 Myr before the final merger (see
Figs. \ref{FigPlaneOfSky} and lower panel in Fig. \ref{FigMergerTimeSequence}).
In the position-velocity diagram (Fig. \ref{FigVcube}) we see three projections in the a) X-Z
(plane-of-the-sky), b) X-Vy  and c) Vy-Z planes. Only gas
particles are shown, where blue and red particles indicate particles
from NGC 4039 and 4038, respectively. 
Comparing the model with observations (overlaid in yellow), we obtain
a good fit with the observed morphology of the
system (left panel in Fig. \ref{FigVcube}). However, the agreement with the observed velocities of the
northern tail could be improved. In this region the data show
overall\linebreak smaller values (right panel in Fig.\ref{FigVcube}).
This could be achieved by slight changes of the orientation of NGC 4039 with
respect to the orbital plane and/or by choosing a slightly different
viewing angle. Also the spatial and kinematical data at the bending
end of the southern tail still need adaptation (middle/right
panel in Fig. \ref{FigVcube}), which may be achieved
by adopting a flat distribution for the initial gas disks.
Our model is similar to the model by Hibbard (2003), but in addition we included gas dynamics, star formation, and feedback in our
simulations. 

\begin{figure}
\centering
\includegraphics[width=80mm, height=40mm]{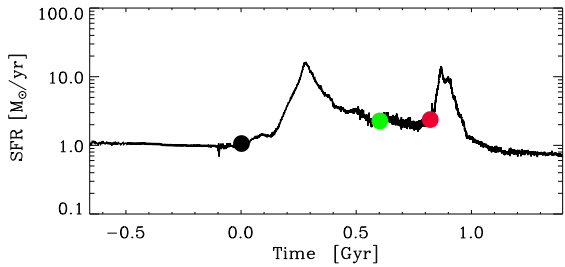}
\caption{Time evolution of the total SFR in the NGC 4038/4939
  model. With filled dots we are indicating the first pericenter ($t=0.00$ Gyr, black), best match ($t=0.60$ Gyr, green)
, and final merging ($t=0.82$ Gyr, red).}
\label{FigTotalSFR}
\end{figure}

In contrast to observations, we do not see enhanced star
formation in the 'overlap' region at the time of 'best fit' as is
observed in the Antennae (see, e.g. Wang et al.~2004). This may be due
to the fact that a significant amount of gas has already been consumed
by the starburst during the first encounter. This will be subject to further
investigations. Fig. \ref{FigTotalSFR} shows the global star formation
rate (SFR) of the simulated Antennae system. The star formation
history in the
two galaxies is dominated by two major starbursts resulting from gas funneled to the
galactic centers after the first ($t = 0.0$
Gyr, black dot) and second passage ($t \approx 0.82$ Gyr, red dot). At
the time of best fit ($t \approx 0.60$ Gyr, green dot) we have a SFR
of $\sim 2.3\, \mathrm{M_\odot\,yr^{-1}}$ which lies below the range of observed
values. Zhang, Fall \& Whitmore~(2001) report a SFR of 4 $< \dot
M_{\ast}/\mathrm{M_\odot\, yr^{-1}} <$
21 derived from the total $\mathrm{H}_\alpha$ flux within the disks of the Antennae 
 either directly (lower value) or with a correction for extinction
 (upper value). Stanford et al.~(1990) give a SFR of $\dot M_{\ast} = 8.8\,\mathrm{M_{\odot} \
yr^{-1}}$ for the total system. Our result is qualitatively consistent
with the low values
of the total SFR of $5.4\, \mathrm{M_{\odot}\,yr^{-1}}$ reported in the Antennae simulations by Mihos, Bothun
\& Richstone (1993).
\section{Summary \& Outlook}
In this work we have presented first steps towards a good numerical model for the Antennae
galaxies. The simulations include star formation and stellar feedback using sub-grid
physics as proposed by Springel \& Hernquist (2003). We are able to
obtain a close match to the observed morphology of the system, while
a perfect fit to the kinematical HI data is still to be
achieved. Our goal is to
investigate in detail the history and distribution of newly formed
stars in the Antennae galaxies, in
particular in the 'overlap' region of the two progenitor disks. Close
comparison to observations then enables us to adjust the star
formation prescriptions used in numerical simulations and figure out
the important physical properties controlling star formation in
interacting galaxies. In this spirit, it is worth examining
alternative formulations for the star formation laws in numerical
simulations, see e.g. Barnes'~(2005) shock-induced star formation
model for the 'Mice' galaxies. There is a number of possible processes
relevant for the enhanced star formation in the Antennae merger,
specifically in the 'overlap' region. One idea is that collisions of
giant molecular clouds (GMCs) during the merger process could trigger
starbursts (Noguchi 1991) in interacting galaxies. As the disk filling
factor of GMCs is rather low ($f < 0.01$), Jog \& Solomon~(1992)
proposed that a hot ionized, high-pressure gas medium, resulting from
fast collisions between HI clouds, could cause a radiative shock
compression in the outer layers of pre-existing GMCs, thereby inducing
bursts of star formation. Elmegreen \& Efremov~(1997) advocate
high-pressure regions, which originate in the large-scale shocks of
interacting galaxies, for being the driver of bound star cluster
formation, while, lately, Renaud et al.~(2008) raised the idea that
compressive tidal modes in galaxy mergers could play an important role
for the formation of globular clusters. It would also be interesting
to study whether one can form candidates for tidal dwarf galaxies in
the tails of the merger as is observed in the southern tidal arm of
the Antennae (Schweizer 1978, Mirabel, Dottori \& Lutz 1992, see also
Wetzstein, Naab \& Burkert 2007).
\acknowledgements
This work was funded by the DFG priority program SPP 1177 ``Witnesses
of Cosmic History: Formation and evolution of galaxies, black holes,
and their environment''. The numerical simulations were performed on
the local SGI-Altix 3700 Bx2, which was partly funded by the Cluster
of Excellence: ``Origin and Structure of the Universe''.


\begin{thebibliography}{}
  \bibitem{} Baldi, A., Raymond, J.C., Fabbiano, G., Zezas, A., Rots,
    A.H., Schweizer, F., King, A.R., Ponman, T.J.: 2006, ApJ 636, 158
  \bibitem{} Barnes, J.,E.: 1988, ApJ 331, 699
  \bibitem{} Barnes, J.,E.: 2004, MNRAS 350, 798
  \bibitem{} Brandl, B.R., Clark, D.M., Eikenberry, S.S., et al.: 2005, ApJ 635, 280
  \bibitem{} Dubinski, J., Mihos, J.C., Hernquist, L.: 1996, ApJ 462, 576
  \bibitem{} Duncan, J.C.: 1923, ApJ 57, 137
  \bibitem{} Elmegreen, B.G., Efremov, Y.N.: 1997, ApJ 480, 235
  \bibitem{} Hernquist, L.: 1990, ApJ 356, 359
  \bibitem{} Hibbard, J.E.: 2003,  BAAS, 35, 1413; and ``http://www.cv.nrao.edu/~jhibbard/n4038/n4038sim/''
  \bibitem{} Hibbard, J.E., van der Hulst, J.M., Barnes, J.E., Rich,
    R.M.: 2001, AJ 122, 2969
  \bibitem{} Jog, C.J., Solomon, P.M.: 1992, ApJ 387, 152
  \bibitem{} Johansson, P.H., Naab, T., Burkert, A.: 2008, arXiv
    0802.0210v1 
  \bibitem{} Katz, N., Weinberg, D.H., Hernquist, L.: 1996, ApJS 105,
    19
  \bibitem{} Mihos, J.C., Bothun, G.D., Richstone, D.O.: 1993, ApJ
    418, 82
  \bibitem{} Mirabel, I.F., Dottori, H., Lutz, D.: 1992, A\&A 256, L19
  \bibitem{} Monaghan, J.J.: 1992, ARA\&A 30, 543
  \bibitem{} Naab, T., Burkert, A.: 2003, ApJ 597, 893
  \bibitem{} Naab, T., Jesseit, R., Burkert, A.: 2006, MNRAS 372, 839
  \bibitem{} Navarro, J.F., Frenk, C.S., White, S.D.M.: 1997, ApJ 490, 493
  \bibitem{} Noguchi, M.: 1991, MNRAS 251, 360
  \bibitem{} Renaud, F., Boily, C.M., Fleck, J.J., Naab, T., Theis,
    C.: 2008, MNRAS, in press
  \bibitem{} Rubin, V.C., Ford, W.K., D'Odorico, S.: 1970, ApJ 160, 801    
  \bibitem{} Saviane, I., Momany, Y., Da Costa, G.S., Rich, R.M.,
    Hibbard, J.E.: 2008, ApJ 678, 179
  \bibitem{} Schweizer, F.: 1978, IAUS 77, 279
  \bibitem{} Schweizer, F., Burns, C.R., Madore, B.F., et al.: 2008, arXiv 0807.3955v1
  \bibitem{} Springel, V.: 2005, MNRAS 364, 1105
  \bibitem{} Springel, V., Hernquist, L.: 2003, MNRAS 339, 289
  \bibitem{} Springel, V., Di Matteo, T., Hernquist, L.: 2005, MNRAS
    361, 776    
  \bibitem{} Stanford, S.A., Sargent, A.I., Sanders, D.B., Scoville,
    N.Z.: 1990, ApJ 349, 492
  \bibitem{} Toomre, A.: 1977, in The Evolution of Galaxies and
    Stellar Populations, ed. B.M. Tinsley \& R.B. Larson (New Haven,:
    Yale Univ. Press), 401
  \bibitem{} Toomre, A. \&  Toomre, J.: 1972, ApJ 178, 623
  \bibitem{} Wang, Z., Fazio, G.G., Ashby, M.L.N., et al.: 2004, ApJS 154, 193
  \bibitem{} Wetzstein, M., Naab, T., Burkert, A.: 2007, MNRAS 375, 805
  \bibitem{} Whitmore, B.C., Zhang, Q., Leitherer, C., Fall, S.M.,
    Schweizer, F., Miller, B.W.: 1999, AJ 118, 1551
  \bibitem{} Zezas, A., Fabbiano, G.: 2002, ApJ 577, 762  
  \bibitem{} Zhang, Q., Fall, S.M., Whitmore, B.C.: 2001, ApJ 561, 727
\end{thebibliography}
\end{document}